# Emergent charge order and unconventional superconductivity in pressurized kagome superconductor CsV$_3$Sb$_5$

Lixuan Zheng[1], Zhimian Wu[1], Ye Yang[1], Linpeng Nie[1], Min Shan[1], Kuanglv Sun[1], Dianwu Song[1], Fanghang Yu[1], Jian Li[1], Dan Zhao[1], Shunjiao Li[1], Baolei Kang[1], Yanbing Zhou[1], Kai Liu[1], Ziji Xiang[1], Jianjun Ying[1], Zhenyu Wang[1,2], Tao Wu[1,2,3*] and Xianhui Chen[1,2,3*]

1. CAS Key Laboratory of Strongly-coupled Quantum Matter Physics, Department of Physics, University of Science and Technology of China, Hefei, Anhui 230026, China

2. CAS Center for Excellence in Superconducting Electronics (CENSE), Shanghai 200050, China

3. Collaborative Innovation Center of Advanced Microstructures, Nanjing University, Nanjing 210093, China

*Correspondence to: wutao@ustc.edu.cn, chenxh@ustc.edu.cn

**The discovery of multiple electronic orders in kagome superconductors AV$_3$Sb$_5$ (A = K, Rb, Cs) provides a promising platform for exploring unprecedented emergent physics [1-9]. Under moderate pressure (< 2.2 GPa), the triple-$Q$ charge density wave (CDW) order is monotonically suppressed by pressure, while the superconductivity displays a two-dome-like behavior, suggesting an unusual interplay between superconductivity and CDW order [10,11]. Given that time-reversal symmetry breaking and electronic nematicity have been revealed inside the triple-$Q$ CDW phase [8,9,12,13], understanding this CDW order and its interplay with superconductivity becomes one of the core questions in AV$_3$Sb$_5$ [3,5,6]. Here, we report the evolution of CDW and superconductivity with pressure in CsV$_3$Sb$_5$ by $^{51}$V nuclear magnetic resonance measurements. An emergent CDW phase, ascribed to a possible stripe-like CDW order with a unidirectional 4a$_0$ modulation, is observed between $P_{c1}$ ~ 0.58 GPa and $P_{c2}$ ~ 2.0 GPa, which explains the two-dome-like superconducting behavior under pressure. Furthermore, the nuclear spin-lattice relaxation measurement reveals evidence for pressure-independent charge fluctuations above the CDW transition temperature and unconventional superconducting pairing above $P_{c2}$. Our results not only shed new light on the interplay of superconductivity and CDW but also reveal novel electronic correlation effects in kagome superconductors AV$_3$Sb$_5$.**

In quantum materials, most notably high-temperature superconductors, the electronic phase diagrams are typically complex with multiple distinct broken-symmetry phases [14,15]. A conventional understanding is to describe these ordering tendencies as "competing order" because one order is usually suppressed by the other with changing composition, pressure or magnetic field. Recently, "intertwined order" has been proposed as an alternative scenario that emphasizes a different perspective by focusing on the cooperative character of different orders [16-18]. A prime example for intertwined orders is the pair-density-wave (PDW) order in cuprate superconductors, which can be entangled with superconductivity and density waves [18]. In this context, electronic nematicity, which is widely observed in cuprates and iron-based superconductors [17] as a vestigial order, is also a particular manifestation of intertwined orders. At the present stage, exploring the underlying physics of intertwined orders in other quantum materials, in particular for superconducting materials, is highly desirable but still rare [19].

Recently, the discovery of vanadium-based kagome metal $A$V$_3$Sb$_5$ ($A$ = K, Rb, Cs) with superconductivity and CDW has generated intense interest, especially for exploring intertwined orders [1,6,8-11]. Earlier theoretical calculations have motivated that, as the filling level close to the van Hove singularities, electron-electron interaction and inherent geometrical frustrations can lead to a large variety of electronic orders in a two-dimensional (2D) kagome lattice [20-23]. Experimentally, scanning tunneling microscopy/spectroscopy (STM/S) experiments have revealed a triple-$Q$ $2a_0 \times 2a_0$ CDW order in $A$V$_3$Sb$_5$ [3-6], and the angle-resolved photoemission spectroscopy (ARPES) experiments have found a momentum-dependent CDW gap [7,24]. Density functional theory (DFT) calculations find unstable phonon modes at both M and L points that seems to favor an electron-phonon driven CDW in $A$V$_3$Sb$_5$ [25]. On the other hand, time-reversal symmetry breaking and electronic nematicity have been revealed inside the triple-$Q$ CDW state [5-9,12,13], implying an unconventional nature of the CDW that stems from a delicate cooperation of electron-phonon interaction, electron-electron interaction and inherent geometrical frustration in $A$V$_3$Sb$_5$ (Fig. 1a) [26-30]. As for the superconducting state, a previous nuclear quadrupole resonance (NQR) experiment observed a pronounced Hebel-Slichter coherent peak just below the superconducting transition temperature ($T_c$) in CsV$_3$Sb$_5$, strongly supporting a conventional $s$-wave pairing [31]. This result is complemented by specific heat and penetration depth measurements. Interestingly, STM/S experiments revealed non-uniform superconductivity in $A$V$_3$Sb$_5$ with a $4a_0/3$ spatial modulation, which has been ascribed to the formation of a $3Q$-PDW state [6]. Furthermore, the superconductivity under moderate pressure ($P <$ 2.2 GPa) displays a puzzling two-dome-like superconducting behavior, while the CDW is monotonically suppressed by increasing pressure [10,11]. All these facts suggest a complicated interplay between superconductivity and CDW in $AV_3Sb_5$. Here, in order to clarify the correlation between superconductivity and CDW, we systematically study the evolution of CDW and superconductivity with pressure in CsV$_3$Sb$_5$ by performing nuclear magnetic resonance (NMR) measurements on $^{51}$V nuclei under hydrostatic pressures up to 2.3 GPa.

**Pressure-dependent superconductivity and CDW order**

First, we revisit the superconducting regime in the $T$-$P$ phase diagram of CsV$_3$Sb$_5$ by utilizing the NMR tank circuit to measure the $P$-dependent superconducting transition. As shown in Fig.1c, the $P$-dependent superconducting transition is monitored by measuring the $T$-dependent relative change of resonating frequency ($-\frac{\Delta f}{f}$) at zero magnetic field (see the Methods for details). Our present result confirms the previously reported two-dome-like behavior of superconductivity with an optimal $T_c$ located at $\sim$ 0.7 GPa and $\sim$ 2.0 GPa, respectively (Fig. 1b) [10,11]; moreover, an unusual superconducting transition broadening is observed in the intermediate pressure range with reduced superconducting temperature (Fig. 4), which is also consistent with earlier transport measurements [10,11].

To understand the two-dome-like behavior of superconductivity, we systematically measured the NMR spectrum of $^{51}$V nuclei at $\sim$ 2 K to study the evolution of CDW order with pressure in the same pressure range. At ambient pressure, the vanadium kagome lattice is reconstructed into a $2a_0 \times 2a_0$ superlattice with two different vanadium sites (Fig. 1e) by a triple-$Q$ CDW order [3-6]. The full NMR spectrum splits into two sets of peaks with the equal integral intensity, suggesting two different structural sites on $^{51}$V nuclei in the CDW state [9,31, 32] (see Methods). For simplicity, we hereafter use the central transition lines to illustrate the $P$- or $T$-dependent evolution of the CDW order. As shown in Fig.1d, the two-peak structure of the NMR spectrum persists up to $P_{c1} \sim$ 0.58 GPa. Once $P$ increases above $P_{c1}$, the NMR spectrum changes remarkably and exhibits a characteristic structure with multiple peaks, suggesting the formation of a new CDW order which will be discussed in detail later. Notably, the new CDW order eventually vanishes above $P_{c2} \sim 2.0$ GPa, where the NMR spectrum restores the single-peak structure of a perfect kagome lattice without any CDW order (Fig.1d and 1g). It should be noted that the $P$-dependent evolution of the CDW state exhibits the typical behavior of a first-order phase transition around $P_{c1}$ and $P_{c2}$, which is implicit in the coexistence of characteristic structures for different phases in NMR spectra (see the Methods for more discussion). On the other hand, the superconducting transition around $P_{c1}$ and $P_{c2}$ also exhibits a typical behavior of first-order phase transition with two superconducting transition temperatures ($T_{c1}$ and $T_{c2}$) in the $T$-dependent $-\frac{\Delta f}{f}$ (see Fig. 1b and extended Fig. 1 for definition of $T_{c1}$ and $T_{c2}$), suggesting a lower $T_c$ in the new CDW phase (Fig. 4).

These results indicate that the $P$-dependent evolution of the CDW perfectly coincides with the two-dome-like superconducting behavior under pressure, suggesting a complex interplay of the new CDW with superconductivity. Interestingly, even the new CDW phase well develops into a pure phase in the intermediate pressure range from 0.9 GPa to 1.7 GPa (Fig. 4 and see more evidence in Extended fig. 1), there still exists an unusual superconducting broadening beyond the picture of first-order phase transition [10, 11]. Whether this is ascribed to fragile superconductivity or superconducting fluctuations as doped La$_2$CuO$_4$ [33] (see Supplementary Information S1 for more details) needs more investigation in the future.

**Multi-step CDW transitions under pressure**

Besides the $P$-dependent evolution of the CDW order at $\sim 2$ K, we also carefully studied the $T$-dependent evolution of the CDW order under different pressures. As shown in Fig. 2, the $T$-dependent NMR spectra were measured under different pressures. Below $P_{c1}$, the NMR spectra split into two main peaks below the CDW transition temperature ($T_{CDW}$) due to the triple-$Q$ CDW order in the kagome plane [9,31,32]. Previous studies have revealed a $\pi$ phase shift of CDW modulation between the neighboring kagome planes, which leads to a three-dimensional (3D) CDW that breaks the rotational symmetry [4,32]. Therefore, in addition to the major splitting, each NMR peak also experiences a secondary splitting just below $T_{CDW}$ due to such an interplane phase shift [9]. In principle, the triple-$Q$ CDW order is described by a multi-component order parameter ($\psi_1$, $\psi_2$, $\psi_3$), where $\psi_i$ (i = 1, 2, 3) represents the component with wave vector $Q_{2a}^i$ [26,29]. When $\psi_1 = \psi_2 = \psi_3$, the triple-$Q$

CDW order preserves the rotational symmetry of the kagome lattice. A previous NMR study on the secondary splitting indicated that, although explicit rotational symmetry breaking with $\psi_1 = \psi_2 \approx \psi_3$ occurs just below $T_{CDW}$, a novel electronic nematicity with $\psi_1 = \psi_2 \neq \psi_3$ is developed below a characteristic temperature ($T_{nem}$); this result is supported by both STM/S and elastoresistivity measurements [9]. In the present work, the $P$-dependence of $T_{nem}$ is monitored by measuring the $T$-dependent linewidth and $1/T_1T$ at various pressures (see the Extended Fig. 2 and Methods). It is found that $T_{nem}$ is almost $P$-independent below $P_{c1}$.

As $P$ is increased just above $P_{c1}$, the $T$-dependent NMR spectra display a two-step transition into the new CDW state. As shown in Fig. 2b, at $P = 0.67$ GPa, the NMR spectra first split into two main peaks below $T_{CDW}$, consistent with a triple-$Q$ CDW state with $\psi_1 = \psi_2 \approx \psi_3$. As the temperature further decreases, the NMR spectrum develops several new peaks below another characteristic temperature (Fig. 2f), indicating the formation of a new CDW order instead of the electronic nematicity. As we have mentioned before, the $P$-dependent evolution from triple-$Q$ CDW to the new CDW displays the typical behavior of a first-order phase transition above $P_{c1}$, which leads to the coexistence of characteristic structures from these two different CDW phases at $P = 0.67$ GPa. When the new CDW state is fully developed with further increasing pressure, the characteristic signature related to the triple-$Q$ CDW (with two main peaks) disappears in the NMR spectrum (Fig. 1d). A single transition to the well-developed new CDW order is clearly manifested in the $T$-dependent NMR spectra at $P = 1.05$ GPa (Fig. 2c and Fig. 2g). With $P$ approaching $P_{c2}$, the NMR spectrum exhibits another first-order phase transition from the new CDW to the kagome phase (Fig. 2d and Fig. 2h). Eventually, the new CDW order completely vanishes at $P_{c2}$, above which no significant change in the NMR spectra can be seen down to the lowest $T$ (Fig. 2e).

An immediate question is how to understand this new CDW order. Since its characteristic spectrum is distinct from that of the original triple-$Q$ CDW, we can confidently rule out the possibility of different types of triple-$Q$ CDW (e.g., superimposed tri-hexagonal and star-of-David modulation along the $c$ axis) [25,26] (see the Methods and Extended Fig. 3 for more analysis). On the other hand, our results indicate that this new CDW phase is evolved from the electronic nematicity inside the triple-$Q$ CDW phase (Fig. 4). In this sense, a stripe-like CDW order, which breaks both rotational and translational symmetries, should be the most promising candidate for the new CDW order (see the Methods for more discussion). This picture is further supported by a quantitative simulation of the NMR spectrum using a unidirectional $4a_0$ modulation (see the Methods and Extended Fig. 3 for more details), although further confirmation is undoubtedly needed. In the following part, the new CDW is named the stripe-like CDW and the corresponding transition temperature is defined as $T_{stripe}$.

**Electronic fluctuations and unconventional superconductivity around $P_{c2}$**

The $T$-dependent nuclear spin-lattice relaxation rate ($1/T_1$) on $^{51}$V nuclei measured under different pressures are shown in Fig. 3. All the CDW transition temperatures determined from the $1/T_1T$ curves are consistent with those from the NMR spectrum (Fig. 3b, 3c and 3d). More importantly, the $1/T_1T$ data taken under different pressures collapse into a universal curve as a function of $T$ above $T_{CDW}$ or $T_{stripe}$, which follows a Curie-Weiss-like behavior (Fig. 3a). In general, $1/T_1T$ measures the dynamic magnetic susceptibility at the NMR frequencies, which depicts the low-energy magnetic correlations. Here, the Curie-Weiss-like behavior of $1/T_1T$ is ascribed to correlation-enhanced electronic fluctuations (see the Methods for more explanation). Usually, quantum critical fluctuations exhibit strong pressure or chemical doping dependence in many correlated electronic systems, such as critical antiferromagnetic/nematic fluctuations in iron-pnictides superconductors [34,35]. Obviously, this is not the case for the observed electronic fluctuations in CsV$_3$Sb$_5$. The pressure-independent Curie-Weiss-like behavior of $1/T_1T$ indicates that the CDW transition only interrupts the development of pressure-independent electronic fluctuations below $T_{CDW}$, which has strong implications for understanding the mechanism of the CDW transition in the AV$_3$Sb$_5$ family.

Considering the absence of local spins in this system [36] and the low-temperature charge order, the observed electronic fluctuations should be ascribed to charge fluctuations instead of spin fluctuations (see the Methods for more discussion). This is also consistent with an abrupt reduction of $1/T_1T$ just after the first-order CDW transition at $T_{CDW}$ (Fig. 3b, 3c and 3d). On the other hand, similar Curie-Weiss-like behavior in $1/T_1T$ is also confirmed by NQR measurement on $^{121/123}$Sb nuclei at zero field (the inset of Fig. 3a and extended Fig. 5), in which the charge fluctuations lead to faster relaxation in the magnetic channel instead of the quadrupole channel. Recent theoretical analysis indicates that the electronically mediated charge order in the AV$_3$Sb$_5$ family preferably breaks time-reversal symmetry due to the appearance of orbital currents [28,37]. In fact, previous zero-field μSR experiments indeed reveal a weak signal for time-reversal symmetry breaking below $T_{CDW}$ [8,12], which is beyond the sensitivity of the NMR technique on magnetic ordering [32] (also see the extended Fig. 6). All these results offer one possibility to understand why charge fluctuations can lead to faster nuclear relaxation in magnetic channels. In previous NQR measurements at ambient pressure, the $1/T_1T$ is almost $T$-independent at low temperatures and exhibits a Hebel-Slichter coherent peak just below $T_c$ [31]; such a peak is usually taken as a smoking gun for conventional Bardeen-Cooper-Schrieffer (BCS) superconductivity [38]. A similar behavior in $1/T_1T$ is revealed by our NQR measurement at low pressure ($P = 0.08$ GPa, see the extended Fig. 5). However, when $P > P_{c2}$, in addition to the Curie-Weiss-like behavior above $T_c$, $1/T_1T$ exhibits an immediate reduction just below $T_c$ instead of showing a Hebel-Slichter coherent peak (see the Methods for more discussion). The absence of the Hebel-Slichter coherent peak has been widely observed in unconventional superconductors, wherein spin fluctuations play a key role [39]. In addition to spin instability, recent experimental works confirm the ubiquitous presence of hidden charge orders and charge fluctuations in cuprate superconductors, highlighting their close connection with unconventional superconductivity [15,40]. In the present work, both remarkable charge fluctuations and the absence of the Hebel-Slichter coherent peak in pressurized CsV$_3$Sb$_5$ strongly indicate a 'clean' platform without local spin fluctuations for studying the correlation between unconventional superconductivity and charge fluctuations (see extended Fig. 7 for quantitative analysis of the correlation between $T_c$ and $1/T_1T$).

**Electronic phase diagram**

As shown in Fig. 4, we summarize the main findings into a pressure-dependent electronic phase diagram, in which a new CDW order is found and ascribed to stripe-like CDW order with a unidirectional 4a$_0$ modulation. Below $P_{c1}$, the leading CDW order remains a triple-$Q$ CDW order but exhibits a novel electronic nematicity below an almost pressure-independent $T_{nem}$. As the pressure increases above $P_{c1}$, a stripe-like CDW order evolves from the electronic nematicity in the triple-$Q$ CDW through a first-order phase transition. We note that just above $P_{c1}$ a two-phase coexisting regime is developed, wherein the system enters the triple-$Q$ CDW state upon cooling and then a coexisting phase with both stripe-like CDW state and nematic triple-$Q$ CDW state. This two-phase coexisting regime terminates at ~ 0.9 GPa, above which there is only a one-step transition into the pure stripe-like CDW order. Eventually, the stripe-like CDW is completely suppressed at $P_{c2}$ through another first-order phase transition, and then the system restores to an undistorted kagome lattice that exhibits significant electronic fluctuations as temperature decreasing. These results indicate that the stripe-like CDW order is tied to a significant suppression of superconductivity between $P_{c1}$ and $P_{c2}$, suggesting a complicated interplay of the stripe-like CDW order with superconductivity.

What is the underlying physics behind the pressure-induced stripe-like CDW order? Theoretical analysis suggests that the unstable phonon modes at both M and L points must play a key role in driving the triple-$Q$ CDW order at ambient pressure [25]. Our DFT calculation excludes significant changes in band structure under moderate pressure; the most notable change induced by pressure is the phonon dispersion (Extended Fig. 4 and Methods). Although the modified phonon dispersion may lead to an adjustment of the triple-$Q$ CDW order, it seems insufficient to account for a stripe-like CDW order [26]. This emphasizes the necessity of the inclusion of electronic correlation. We note that the pressure-independent electronic fluctuations revealed by $1/T_1T$ suggest an inherent electronic instability in $CsV_3Sb_5$, which is also supported by different theoretical calculations on 2D kagome lattices [22,23,26-30]. Considering all these facts, the pressure-induced stripe-like CDW should stem from an intriguing interplay of electron-phonon interaction and electron-electron interaction in the kagome lattice, which calls for further theoretical investigations.

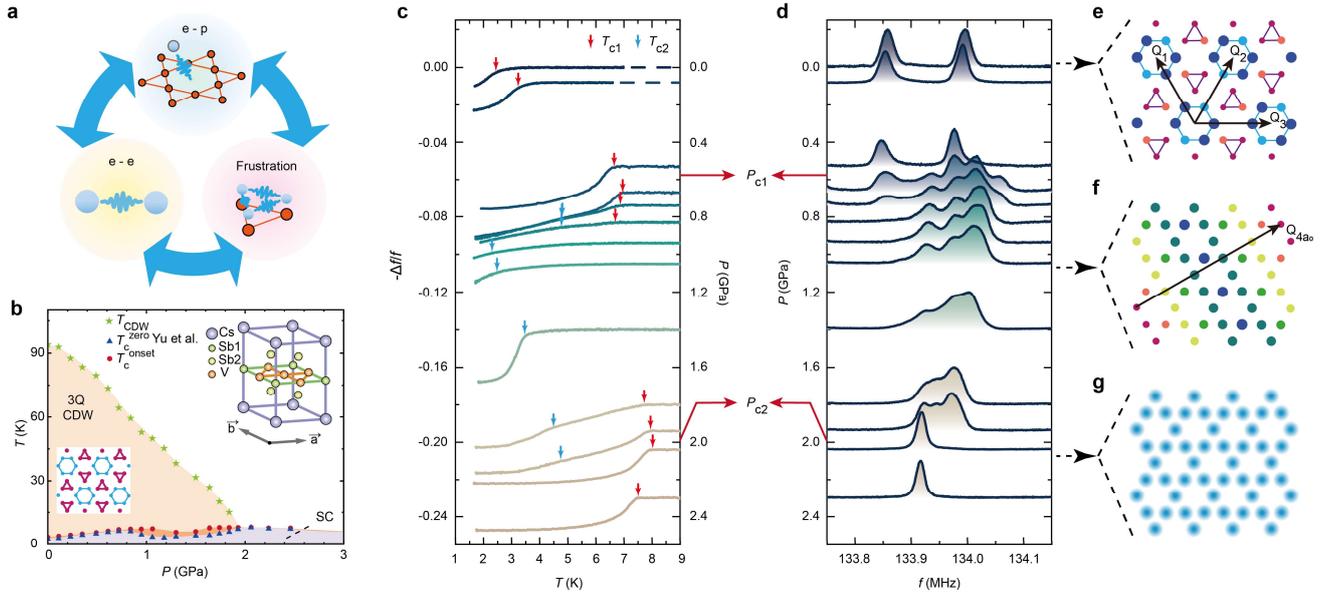

**Figure 1 | Pressure-dependent evolution of superconductivity and CDW state in CsV$_3$Sb$_5$. a,** A cartoon illustration of the interplay among electron-phonon, electron-electron and geometric frustration, which plays a key role in the competing electronic orders in CsV$_3$Sb$_5$. **b,** Pressure-dependent electronic phase diagram obtained in a previous study [11]. The inset at the upper right corner is the unit cell of CsV$_3$Sb$_5$. The inset at the lower left corner is a schematic plot of the 2a$_0$ × 2a$_0$ superlattice in real space. **c,** The temperature-dependent resonating frequency of the NMR tank circuit under different pressures. The red and blue arrows label two superconducting transition temperatures $T_{c1}$ and $T_{c2}$, respectively. The definition is provided in the Methods for details. In each curve, the zero value of - Δ $f/f$ is shifted to the value of applied pressures on the right-hand longitudinal axis. **d,** Pressure-dependent evolution of $^{51}$V NMR spectra at 2 K. Only the central line is plotted here. The external magnetic field is along the $c$ axis with a magnitude of 11.9 T. The background of each spectrum is also shifted to the value of applied pressures on the left-hand longitudinal axis. **e, f, g,** The schematic plots of electronic states in real space under different pressures. **e,** triple-$Q$ CDW with $\psi_1$ = $\psi_2 \neq \psi_3$ below $P_{c1}$ ~ 0.58 GPa; **f,** stripe-like CDW with a unidirectional 4a$_0$ modulation between $P_{c1}$ and $P_{c2}$; **g,** the normal metallic phase above $P_{c2}$ ~ 2.0 GPa. All data in this figure were collected on sample B which is carefully stacked along the $c$ axis. Similar results are also repeated in sample A (see Supplementary Information S2).

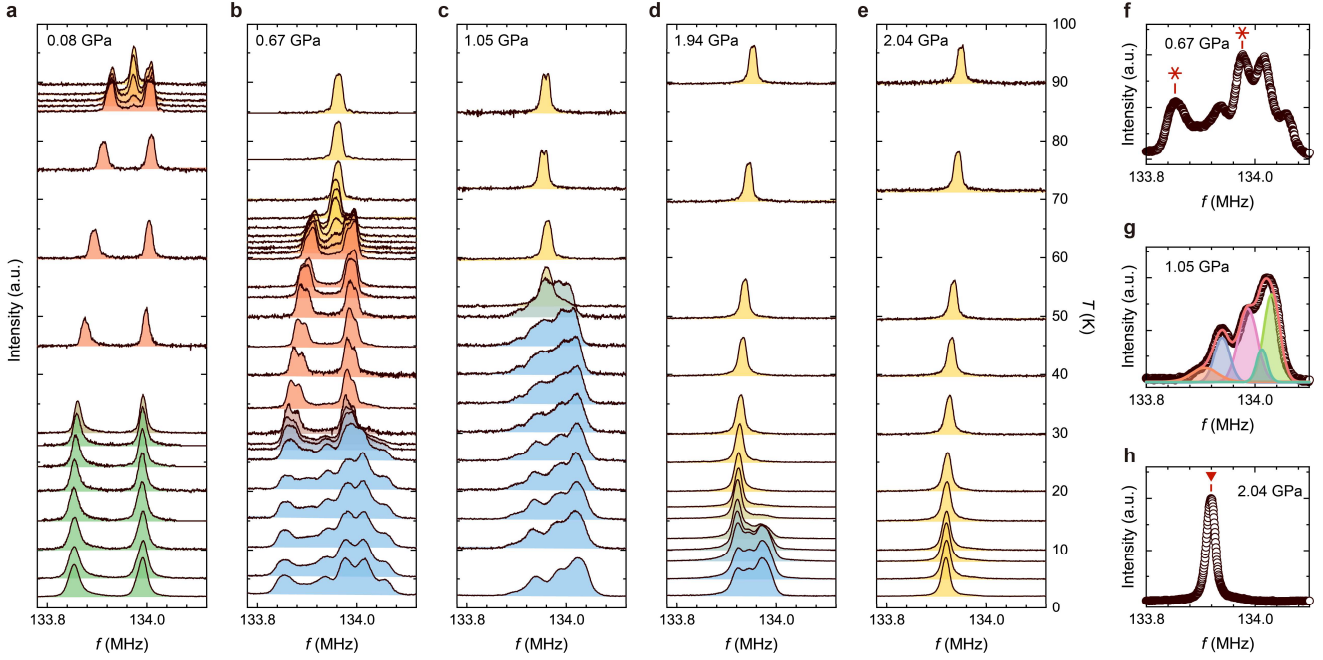

**Figure 2 | $^{51}$V NMR evidence for the pressure-dependent multiple CDW transition. a-e,** The temperature-dependent $^{51}$V NMR central lines under different pressures. The external magnetic field is along the $c$ axis with $\mu_0H = 11.9$ T. Yellow, orange, green and blue shaded areas represent the kagome phase, triple-Q CDW with $\psi_1 = \psi_2 \approx \psi_3$, triple-Q CDW with $\psi_1 = \psi_2 \neq \psi_3$ (or electronic nematicity) and stripe-like CDW, respectively. Due to a first-order phase transition, there is a temperature range exhibiting the coexistence of high-temperature and low-temperature phases. Here, the $T_{CDW}$ is determined when both high-temperature and low-temperature phases have the same peak height. The definition of $T_{nem}$ and $T_{stripe}$ is introduced in the Methods and Extended Figs. 2 and 8. It should be noted that, due to a pressure-induced misalignment among stacking sample B (see Supplementary Information S3), a tiny splitting is observed in the NMR spectrum above $T_{CDW}$ or $T_{stripe}$ for $P > 0.08$ GPa. Such a misalignment is less visible for sample A under pressure (see Supplementary Information S4). The longitudinal axis on the right side shows the temperatures of each spectrum. **f-h,** Representative $^{51}$V NMR central line at 2 K for three different pressure regions. **f,** NMR spectrum with both characteristic peaks for triple-$Q$ CDW and stripe-like CDW at $P = 0.67$ GPa. Red asterisks label the characteristic peaks for triple-$Q$ CDW. **g,** NMR spectrum for a fully developed stripe-like CDW at $P = 1.05$ GPa. The colorful curves are fitting results obtained by considering a unidirectional $4a_0$ modulation (see the Methods for details). **h,** NMR spectrum at $P = 2.04$ GPa. The red triangle labels the characteristic NMR peak for an undistorted kagome lattice for $P > P_{c2}$. All data in this figure were collected in sample B.

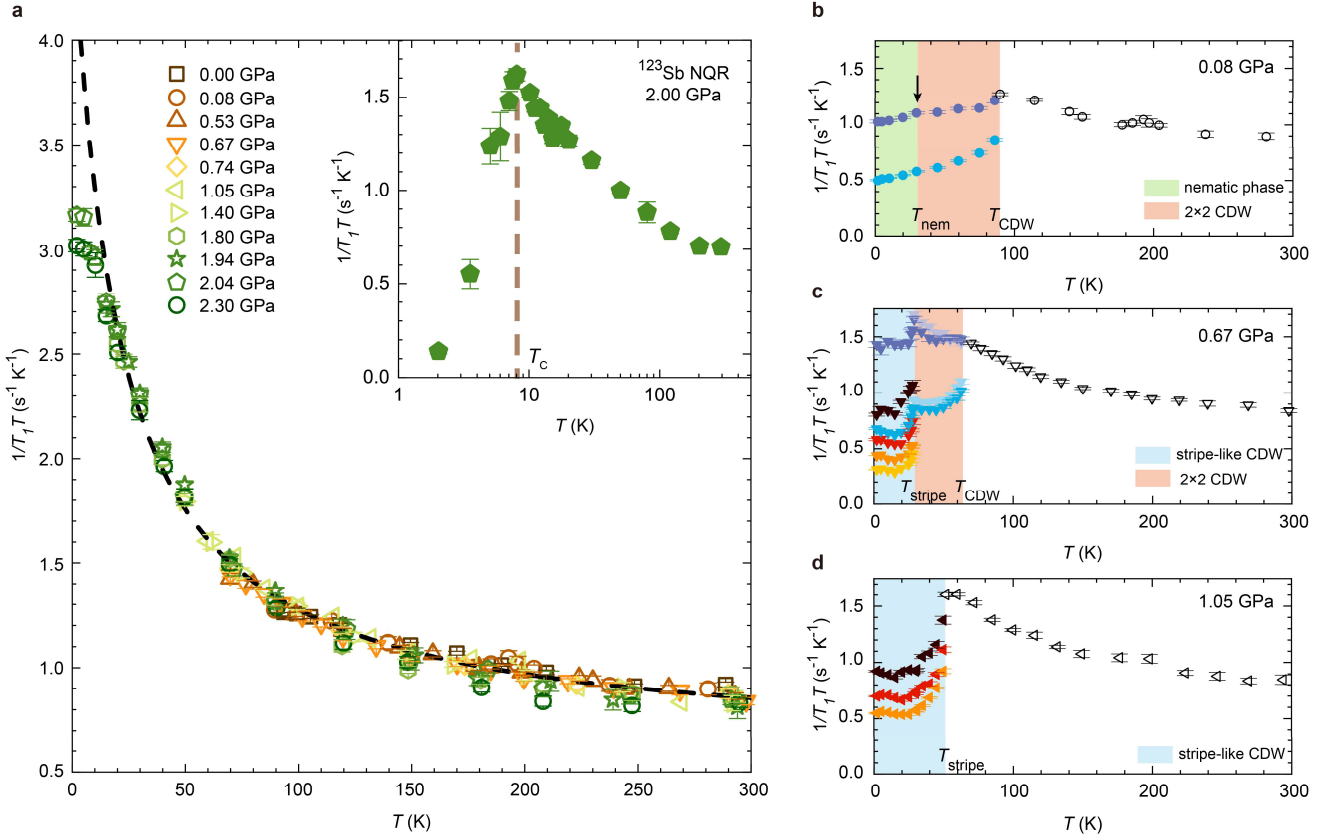

**Figure 3 | Evidence for multiple CDW transitions and persistent charge fluctuations from nuclear spin-lattice relaxation. a**, Temperature-dependent $1/T_1T$ of $^{51}$V NMR above CDW transitions under different pressures. All data points collapse into a single Curie-Weiss-type curve with a slight deviation below ~ 15 K. The black dashed line is a fitting curve of the Curie-Weiss formula with Weiss temperature $\theta$ ~ -21 K. If considering a possible quantum criticality at $P_{c2}$, the deviation below ~ 15 K might be related to the quenched disorder effect on critical fluctuations. The inset shows the temperature-dependent $1/T_1T$ by $^{123}$Sb NQR measurement at $P$ = 2.00 GPa, indicating that the Curie-Weiss-like behavior in $1/T_1T$ is persistent until the superconducting transition. **b-d**, Typical temperature-dependent $1/T_1T$ for three different pressure regions. Colored areas label different CDW phases. A kink feature representing the change of order parameters from $\psi_1 = \psi_2 \approx \psi_3$ to $\psi_1 = \psi_2 \neq \psi_3$ in the triple-$Q$ CDW is marked by black arrow in **b**. The signature for the formation of stripe-like CDW can be clearly resolved in $1/T_1T$ curves under 0.67 GPa in **c** and 1.05 GPa in **d**. The details about the measurement and analysis of $1/T_1T$ are shown in the Methods. The different values of $1/T_1T$ shown with different colors are extracted from different frequency integral regions (see details in Extended Fig. 9). Except for the NQR data in the inset figure, which were collected in sample A, the other data were collected in sample B.

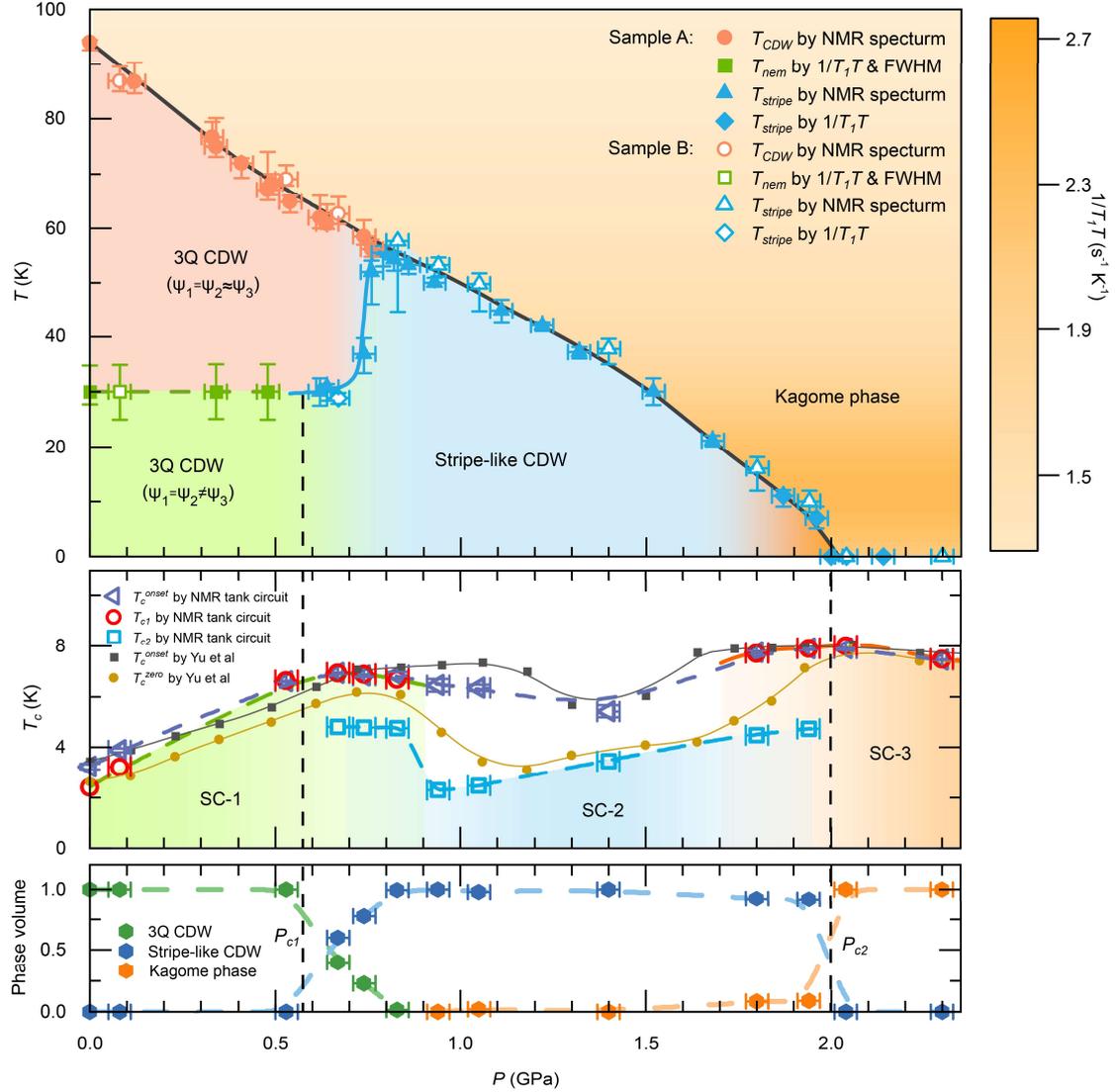

**Figure 4 | Pressure-dependent electronic phase diagram of CsV₃Sb₅.** Three different CDW phases were determined by $^{51}$V NMR measurements. The details for the determination of the CDW phase boundary, $T_c^{onset}$, $T_{c1}$ and $T_{c2}$ are shown in the Methods. The three different CDW phases are shown with different colors. $P_{c1}$ and $P_{c2}$ are defined as the onset pressures of the phase transition toward stripe-like CDW. Due to the first-order quantum phase transition around $P_{c1}$ and $P_{c2}$, we use the mixed color to represent the coexistence of different electronic phases in this region. The temperature-dependent $1/T_1T$ above CDW transitions measured by NMR are plotted as a color background in the phase diagram, indicating pressure-independent charge fluctuations above CDW transitions. The solid/open points represent data collected from sample A/B, respectively. The superconducting temperature defined by the temperature-dependent resonating frequency of the NMR tank circuit ($T_{c1}$, $T_{c2}$ and $T_c^{onset}$) is consistent with that obtained by transport [11] which is also plotted in the phase diagram. The figure in the bottom of the plot is the pressure-dependent evolution of the phase volume for each electronic phase. The definition of phase volume is shown in the Methods and Extended Fig. 10.

## Method

**Sample growth and characterization:** All samples used in the present experiments were synthesized by a self-flux method as previously reported [11]. Sample A (ten pieces of single crystals) and sample B (seven pieces of single crystals) were picked up from two different batches (see Supplementary Information S5). Ten/seven pieces of single crystals for the two samples (A and B) were carefully stacked together along the $c$ axis. Sample C is from the same batch as sample B. The samples picked up from the same batch have nearly the same CDW transition temperatures and superconducting transition temperatures, which is verified by bulk magnetic susceptibility measurements. The bulk magnetic susceptibilities were measured by using a superconducting quantum interference device (Quantum Design MPMS-5).

**NMR and NQR measurements under pressure**: A commercial NMR spectrometer from Thamway Co. Ltd. is used for NMR and NQR measurements. An NMR-quality magnet from Oxford Instruments offers a uniform magnetic field up to 12 Tesla. The external magnetic field at the sample's position is calibrated by $^{63}$Cu NMR of the NMR coil. All NMR spectra were measured by the standard spin echo method with a fast Fourier transform (FFT) sum. All NMR measurements were performed under an external magnetic field of 11.9 T along the $c$ axis. The nuclear spin-lattice relaxation time ($T_1$) of $^{51}$V nuclei was measured by the inverse recovery method and then the recovery of the nuclear magnetization M(t) was fitted by a function with $1 - \frac{M(t)}{M(\infty)} = I_0 \left( \frac{1}{84} \times \exp\left(-\left(\frac{t}{T_1}\right)^{\beta}\right) + \frac{3}{44} \times \exp\left(-\left(\frac{6t}{T_1}\right)^{\beta}\right) + \frac{75}{364} \times \exp\left(-\left(\frac{15t}{T_1}\right)^{\beta}\right) + \frac{1225}{1716} \times \exp\left(-\left(\frac{28t}{T_1}\right)^{\beta}\right) \right)$, in which error bars are determined by the least-square method. The $T_1$ value from $^{123}$Sb NQR was measured on the $(\pm\frac{7}{2} \leftrightarrow \pm\frac{5}{2})$ transition line and the corresponding recovery of the nuclear magnetization M(t) was fitted by a function with $1 - \frac{M(t)}{M(\infty)} = I_0 \left( 0.14 \times \exp\left(-\left(\frac{21t}{T_1}\right)^{\beta}\right) + 0.65 \times \exp\left(-\left(\frac{10t}{T_1}\right)^{\beta}\right) + 0.21 \times \exp\left(-\left(\frac{3t}{T_1}\right)^{\beta}\right) \right)$. For the case of the $^{121}$Sb NQR on the $(\pm\frac{3}{2} \leftrightarrow \pm\frac{1}{2})$ transition line, the recovery function is $1 - \frac{M(t)}{M(\infty)} = I_0 \left( 0.893 \times \exp\left(-\left(\frac{10t}{T_1}\right)^{\beta}\right) + 0.107 \times \exp\left(-\left(\frac{3t}{T_1}\right)^{\beta}\right) \right)$. The typical recovery curves are shown in Supplementary Information S6 and the temperature and pressure dependence of parameter $\beta$ in the $T_1$ fitting is shown in Supplementary Information S7.

**Hydrostatic pressure and calibration:** The hydrostatic pressure is realized by utilizing a self-clamped BeCu piston-cylinder cell. The pressure transmitting medium is Daphne 7373 oil. The pressure was calibrated through the NQR frequency of $Cu_2O$ powder. The relation between the pressure and the NQR frequency of $Cu_2O$ powder at different temperatures and pressures has been determined in ref [41]. The Daphne 7373 oil is liquid at room temperature but solidifies as cooling. The pressure gradually changes when approaching the freezing temperature but undergoes little change after solidification. The temperature-dependent pressure with Daphne 7373 oil as the pressure transmitting medium has been studied in ref [42,43]. The corresponding relation between pressure at 2 K and pressure at 300 K is plotted in Fig. S8a of Supplementary Information S8 (red points) and fitted by a linear function:

$$P(2\text{K}) = 1.11 \times P(300\text{K}) - 0.334.$$

The unit in the formula is GPa. This relation was also checked in our pressure cell up to approximately 2 GPa at 2 K (black points in Fig. S8a of Supplementary Information S8). The temperature dependence of pressure was also checked under several pressures as shown in Fig. S8b of Supplementary Information S8. The pressure change below 100 K is less than 0.02 GPa at pressures below 1 GPa. Above 1 GPa, the pressure change below 50 K is less than 0.01 GPa. In the present work, the pressures were first calibrated at 300 K and then the pressures at 2 K were corrected by utilizing the above formula. Finally, all pressures shown in this work are calibrated pressures at 2 K.

**Pressure-dependent superconducting transition:** The NMR tank circuit can be used to detect superconducting transition by measuring the change in the resonant frequency of the NMR tank circuit [44,45]. The NMR coil is one part of the resonant $LC$ tank circuit. When the inserted sample enters the superconducting state, the diamagnetic shielding effect reduces the inductance of the NMR coil and causes the resonant frequency of the NMR tank circuit to increase. The inductance of the NMR coil with a cylindrical superconductor inside can be calculated as: $L = \frac{gN^2}{l}(\mu_0 S_{res} + \mu_{sam} S_{sam}) = \frac{gN^2}{l}\pi[\mu_0(R^2 - r^2) + \mu_{sam}(2r\lambda - \lambda^2)]$, where $g$ is the geometric factor of NMR coil, $N$ is the number of the NMR coil turns, $l$ is the length of the NMR coil, $\mu_0$ is the vacuum permeability, $S_{res}$ is the residual cross-section area of the NMR coil after subtracting the sample's cross-section area, $R$ is the radius of the NMR coil, $\mu_{sam}$ is the average permeability of the sample's normal state, $S_{sam}$ is the cross-section area of the sample, $r$ is the sample's radius and $\lambda$ is the sample's penetration depth. As temperatures are below $T_c$, $\lambda$ is affected by the diamagnetic shielding effect, which leads to a corresponding change in the resonating frequency $f$ of the NMR tank circuit. Then, the change in inductance by diamagnetic shielding can be calculated as $\Delta L = \frac{gN^2}{l}\pi[(2r\lambda - \lambda^2)\Delta\mu_{sam} + 2(r - \lambda)\mu_{sam}\Delta\lambda]$. Considering that $\Delta\mu_{sam}$ is negligible compared to $\Delta\lambda$ during the superconducting transition and $r \gg \lambda$, $\Delta L$ can be simplified as $\Delta L \cong 2r\mu_{sam}\Delta\lambda$. In the resonant $LC$ tank circuit where the resonant frequency of NMR tank circuit $f = \frac{1}{2\pi\sqrt{LC}}$, the relation between $\Delta f$ and $\Delta L$ is $\frac{\Delta f}{f} = -\frac{\Delta L}{2L}$. Finally, $\Delta f$ is obtained by replacing $\Delta L$ with $\Delta\lambda$: $-\frac{\Delta f}{f} = \frac{r\mu_{sam}}{L}\Delta\lambda \propto \Delta\lambda$. A linearly temperature-dependent background has already been subtracted from $-\frac{\Delta f}{f}$, as shown in Fig. 1c. $T_c^{onset}$ is defined as the starting temperature where the average value of $-\Delta f/f$ deviates from zero. The error bar of $T_c^{onset}$ is defined by the uncertainty due to the background noise of $-\Delta f/f$. $T_{c1}$ and $T_{c2}$ are defined as the superconducting temperature of the 3Q CDW/Kagome phase and stripe-like CDW phase, which are determined by the dip temperature in the second derivation curves of $-\Delta f/f$. The second derivation curves are plotted in Extended Fig. 1. Below $P_{c1}$ or above $P_{c2}$, $T_{c1}$ is very close to $T_c^{onset}$, suggesting a uniform superconducting transition. Between $P_{c1}$ and $P_{c2}$, $T_{c2}$ is much lower than $T_c^{onset}$, suggesting an unusual superconducting broadening effect. The superconducting phase diagram is shown in Fig. 4 and Supplementary Information S9 for sample B and A, respectively.

**First-order phase transition around $P_{c1}$ and $P_{c2}$:** The NMR spectra around $P_{c1}$ and $P_{c2}$ exhibit the characteristic structure for the coexistence of two distinct phases, suggesting a first-order-like quantum phase transition [46]. Usually, the first-order phase transition would lead to an abrupt change in physical properties (including superconductivity) in different phases [47]. Here, the superconducting transition indeed displays a two-step feature around $P_{c1}$ and $P_{c2}$, supporting the first-order phase transition. To quantitatively understand the first-order quantum phase transition at $P_{c1}$ and $P_{c2}$, the phase volume of each phase is analyzed by utilizing the NMR spectrum. As shown in Extended Fig. 10, the change in phase volume can be characterized by the relative intensity of characteristic peaks belonging to each phase. Here, two characteristic peaks were used to determine the phase volume. One is the lowest-frequency peak in the NMR spectrum below ~ 0.9 GPa, which comes from the triple-$Q$

CDW phase. This peak gradually disappears through the first-order quantum phase transition at $P_{c1}$. The phase volume of the triple-$Q$ CDW phase can be estimated by integrating the spectrum intensity of the lowest-frequency peak (for details, see the caption of Extended Fig. 10). The other characteristic peak is the highest-frequency peak in the NMR spectrum above ~ 0.9 GPa, which comes from the stripe-like CDW phase. This peak gradually disappears through the first-order quantum phase transition at $P_{c2}$. The phase volume of the stripe-like CDW phase can be estimated by integrating the spectrum intensity of the highest-frequency peak (for details, see the caption of Extended Fig. 10). By analyzing the relative integrated intensity of these two characteristic peaks, the complete evolution of the first-order quantum phase transition was quantitatively revealed, as shown in Extended Fig. 10. The two samples show similar pressure-driven first-order quantum phase transitions at $P_{c1}$ and $P_{c2}$. In addition, a thermal hysteresis between the triple-$Q$ CDW and stripe-like CDW phases was also revealed in the NMR spectrum at 0.68 GPa (see Supplementary Information S10), further supporting the first-order phase transition between different electronic phases. In fact, the coexistence of the triple-$Q$ CDW vector and stripe-like CDW vector in momentum space has been observed in a previous STM/S experiment on an Sb terminated surface [5]. In that case, a two-step CDW transition was also observed in a temperature-dependent STM/S experiment [5]. All these features are similar to what we observed for the pressure-induced stripe-like CDW. In addition, since Ginzburg-Landau (GL) analysis suggests that incommensurate CDW always appears before commensurate CDW as $T$ decreases [48], the possibility of an incommensurate CDW order to explain the new CDW and above crossover behavior is less likely by considering a two-step CDW transition under $P$ = 0.67 GPa (Fig. 2b).

**Analysis of pressure-dependent NMR spectra with different CDW models:** There are multiple CDW phases in the pressure-dependent electronic phase diagram of CsV$_3$Sb$_5$ as shown in Fig. 4. These different CDW phases can be identified by $^{51}$V NMR spectroscopy. In general, a specific CDW phase could constrain the number of peaks and their intensity ratio in the NMR spectrum, which can be used to identify different CDW models by fitting the NMR spectrum. In Supplementary Information S11, four full $^{51}$V NMR spectra under different pressures at 2 K are plotted. The NMR spectrum under 0.08 GPa reveals two different $^{51}$V sites with equal intensity. This is consistent with a well-established triple-$Q$ CDW order with a 2a$_0$ × 2a$_0$ superlattice [31 ,32,49]. Above $P_{c1}$, the NMR spectrum is significantly changed, suggesting a new CDW order different from the triple-$Q$ CDW order. Based on a previous

STM experiment and theoretical analysis, we selected several candidates for possible CDW modulation, including a unidirectional 1Q CDW with periods of $4/3a_0$ or $2a_0$ or $4a_0$ and a $3Q$ CDW with periods of $4/3a_0$ or $2a_0$ (superimposed tri-hexagonal and star-of-David modulation). We considered the $1Q$ CDW model first. The $1Q$ $2a_0$ CDW is easily ruled out because the corresponding intensity ratio of 1:1:4 does not fit the NMR spectrum. The left two kinds of $1Q$ CDW models with $1Q_{4a_0}$ and $1Q_{4/3a_0}$ cannot be distinguished by fitting the NMR spectrum because their lowest common multiple is $4a_0$. In Extended Fig. 3b, the NMR spectrum is fitted by $1Q_{4a_0}$ modulation with an intensity ratio of 1:1:2:4:4. The fitting result can well explain all features in the observed NMR spectrum. Moreover, the $1Q_{4a_0}$ CDW order has been observed by STM/S experiments on Sb terminated surfaces [4-6], we believe that the $1Q_{4a_0}$ CDW model is more realistic than the $1Q_{4/3a_0}$ CDW model. We also considered the possible $3Q$ CDW orders. The $3Q_{4/3a_0}$ CDW can be ruled out because it comes from Fermi surface reconstruction by a $2a_0 \times 2a_0$ superlattice [50], which is not our situation observed at 1.05 GPa. The possible $3Q_{2a_0}$ CDW with superimposed tri-hexagonal and star-of-David modulation can also be ruled out. Since the interlayer coupling only leads to a secondary effect on the $^{51}$V NMR spectrum [9,49], we would expect two groups of double peaks for the superimposed tri-hexagonal and star-of-David modulation, in which one double peak should be the same as the triple-$Q$ CDW below $P_{c1}$. In Extended Fig. 3e, the observed NMR spectrum at 1.05 GPa obviously cannot be fitted by this kind of modulation. In conclusion, we believe that the most likely CDW modulation to account for our present NMR result is the stripe-like CDW order with a unidirectional $4a_0$ modulation. It should be noted that although a stripe-like CDW order with unidirectional $4a_0$ modulation has been observed by previous STM experiments at ambient pressure on Sb-terminated surfaces of $CsV_3Sb_5$ and $RbV_3Sb_5$ at low temperatures [4-6,51], no evidence for stripe-like CDW order has been found with bulk measurements thus far [1,9,30, 32,49]; whether such stripe-like CDW order is an intrinsic nature for $AV_3Sb_5$ is under intense debate [4-6 52,53]. One recent first-principles calculation work suggests that the formation of stripe-like CDW on an Sb-terminated surface might be sensitive to surface reconstruction [53]. The new CDW phase observed in the bulk here might be attributed to a stripe-like CDW order that is stabilized by pressure (Fig. 1f).

**Determination of the phase boundary between different CDWs:** In addition to the CDW transition to triple-$Q$ CDW, there are two phase boundaries for different CDWs in the pressure-dependent electronic phase diagram in Fig. 4. One is the phase transition to electronic nematicity inside the triple-$Q$ CDW. The rapidly increasing linewidth of $^{51}$V NMR at approximately 35 K has been confirmed as evidence of electronic nematicity at ambient pressure [9]. A similar rapidly increasing linewidth with almost the same starting temperature is revealed below $P_{c1}$ in sample group B (see extended Fig. 2e-h). In addition, the nematic transition also leads to a sudden drop in $1/T_1T$ below $T_{nem}$. The nearly unchanged $T_{nem}$ determined by FWHM and $1/T_1T$ is quite consistent. Above $P_{c1}$, the $^{51}$V NMR spectrum at 2 K changes from a double-peak due to triple-$Q$ CDW to a multiple-peak due to stripe-like CDW. It should be noted that around $P_{c1}$, the value of $T_{nem}$ is equal to the value of $T_{stripe}$, which suggests that the electronic nematicity is a vestigial order of the stripe-like CDW order. Another phase boundary is from the triple-$Q$ CDW to the stripe-like CDW. As shown in Extended Fig. 8**a-e**, the spectrum weight of stripe-like CDW on a selected frequency range ($A_1$) is continuously increased. The $T_{stripe}$ is determined by the temperature-dependent spectrum weight. Here, we defined a renormalized intensity ratio by integrating another spectral weight on the peak belonging to triple-Q CDW ($A_2$). The renormalized intensity ratio is expressed as $A_1/(A_1+A_2)$ and is shown in Extended Fig. 8**f-j**. The spectral integration on both triple-Q CDW and stripe-like CDW is performed on the corresponding peak frequency with a width of 10 kHz. The characteristic peak position is labelled in Extended Fig. 8**a-e** by black or blue bars, which correspond to triple-$Q$ CDW and stripe-like CDW, respectively. A slow enhancement of the renormalized intensity ratio is possibly due to the short-range ordering induced by disorder from 0.53 to 0.67 GPa.

**Analysis of Curie-Weiss-like behavior in $1/T_1T$:** In a weakly interacting Fermi liquid (FL), $1/T_1T$ reflects the density of states at the Fermi level ($\sim D(E_F)^2$) and is expected to follow a standard Korringa relation with the Knight shift ($\sim D(E_F)$). Following the FL picture, the $P$-independent $1/T_1T$ above $T_{CDW}$ or $T_{stripe}$ speaks against remarkable change of electronic structure under pressure (see extended Fig. 4). When the stripe-like CDW is completely suppressed above $P > P_{c2}$, $1/T_1T$ exhibits a continuous increase down to $T_c$ following a Curie-Weiss-like behavior (see the caption of Fig. 3). Since the ARPES experiment reveals multiple vHSs in the vicinity of the Fermi level [2], a strongly $T$-dependent $D(E_F)$ might be responsible for the increase in $1/T_1T$. Alternatively, when the interaction between quasiparticles is turned on, the electronic instability driven by interactions gives rise to electronic fluctuations that would also strongly affect $1/T_1T$ through the dynamic susceptibility (Im$\chi_q(\omega)$) [54]. In such a scenario, the Korringa relation between the $1/T_1T$ and Knight shifts should be broken down. This is indeed the case in pressurized CsV$_3$Sb$_5$: the Korringa plot shows a clear deviation from a standard linear behavior (see Supplementary Information S12), supporting the significant contribution of electronic fluctuations in $1/T_1T$. Therefore, we conclude that the Curie-Weiss-like behavior in $1/T_1T$ is beyond the noninteracting FL description with multiple vHSs [2,55,56] and is linked to correlation-enhanced electronic fluctuations.

**The absence of the Hebel-Slichter peak at $P_{c2}$:** In principle, the absence of the Hebel-Slichter peak in a conventional $s$-wave superconductor can originate from anisotropic or multiple superconducting gaps, a large nuclear electric quadrupole moment and phonon damping in the strong coupling regime [57]. First, the possibility of a large nuclear electric quadrupole moment can be excluded because the nuclear spin-lattice relaxation in the NQR measurement is dominated by magnetic relaxation in our case (see the caption of Extended Fig. 5). Second, although CsV$_3$Sb$_5$ exhibits two quite different superconducting gaps [58], the Hebel-Slichter peak has been observed at both ambient pressure and 0.08 GPa. We think that the possibility of anisotropic or multiple superconducting gaps should be unlikely at $P_{c2}$. Third, the possibility of phonon damping in the strong coupling regime is also unlikely since the drop of $1/T_1T$ below $T_c$ is not fast enough to satisfy a strong coupling behavior.

**The possibility of spin fluctuations:** In the main text, although the Curie-Weiss-like behavior in $1/T_1T$ has been ascribed to CDW fluctuations, the possibility of spin fluctuations still needs more discussion. Previous theoretical work on the kagome lattice has revealed a rich electronic phase diagram in a wide parameter space, in which spin-density-wave order is one of the possible ground states in parameter space [21-23,29]. However, recent dynamical mean-field theory (DMFT) calculations on AV$_3$Sb$_5$ suggest Pauli-like paramagnetism and the absence of local moments, which is consistent with recent experimental results [59]. This result also indicates that the local correlation strength is not decisive and that nonlocal electronic correlation might be crucial in this kagome system, which has been demonstrated in recent theoretical calculations [60]. In this sense, the spin fluctuations are less likely to account for the divergent behavior in $1/T_1T$.

**First-principles calculation:** We performed the density functional theory calculation as implemented in the Vienna ab initio Simulation Package (VASP) [61] within the framework of generalized gradient approximation [62]. The Crimme (DFT-D3) method [63] is employed to describe the van der Waals interactions. The kinetic energy cutoff is set to 350 eV for the plane wave basis, and the Brillouin zone is sampled using $16 \times 16 \times 9$ Monkhorst-Pack grids. Spin orbital coupling (SOC) is included in the electronic structure calculations. The Hellmann-Feynman forces on each atom were relaxed to less than 0.001 eV/Å. The phonon dispersions are obtained by using the PHONOPY code [64]. The lattice parameter used for calculation is shown in Supplementary Information S13.

## Acknowledgments

We thank the valuable discussion with K. Jiang, J. F. He and Y. L. Wang. This work is supported by the National Natural Science Foundation of China (Grants No. 11888101, 12034004, 12074364), the National Key R&D Program of the MOST of China (Grants No. 2017YFA0303000), the Strategic Priority Research Program of Chinese Academy of Sciences (Grant No. XDB25000000), the Anhui Initiative in Quantum Information Technologies (Grant No. AHY160000).

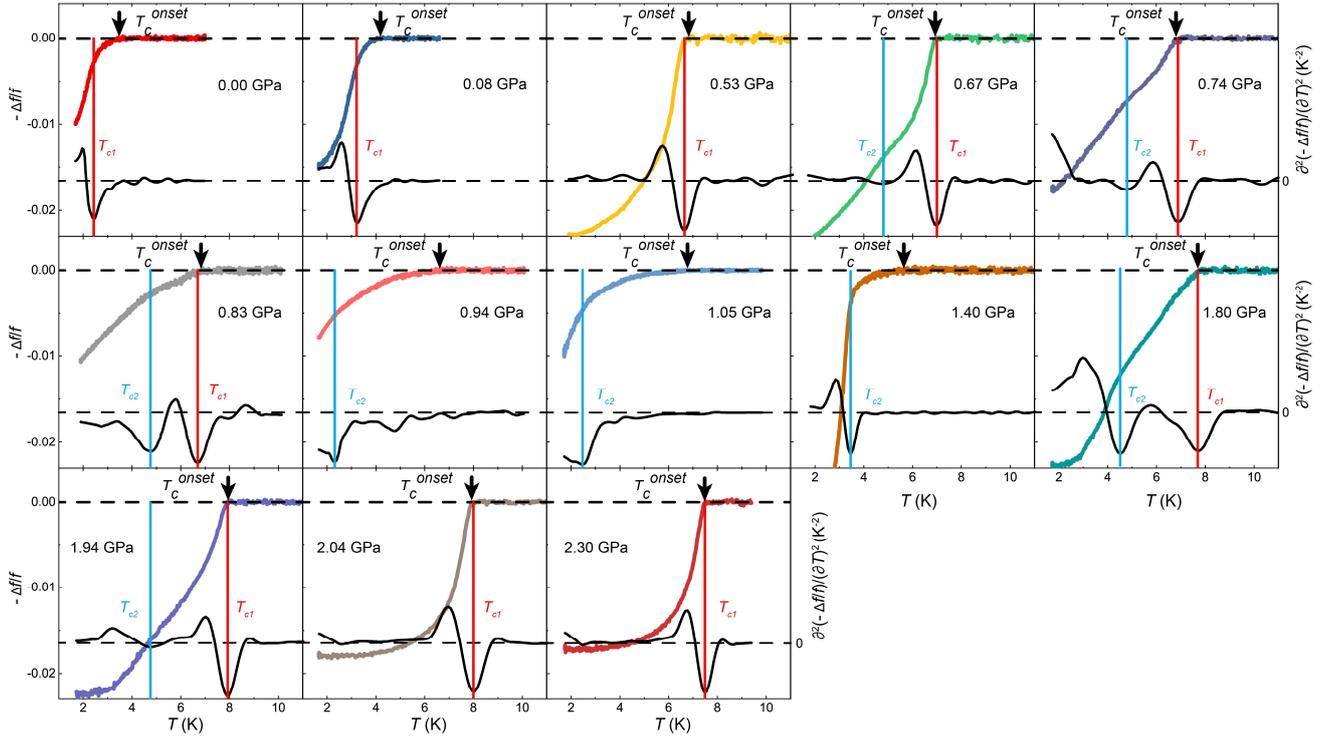

**Extended data Fig. 1 | Determination of $T_c^{onset}$, $T_{c1}$ and $T_{c2}$ from the resonant frequency of the NMR tank circuit.** $T_c^{onset}$ is defined as the temperature where the average value deviates from zero. $T_{c1}$ and $T_{c2}$ are defined by the valley of the second derivation of $-\Delta f/f$. $T_c^{onset}$ is labelled by a black arrow. $T_{c1}$ and $T_{c2}$ are labelled by red and blue curves, respectively.

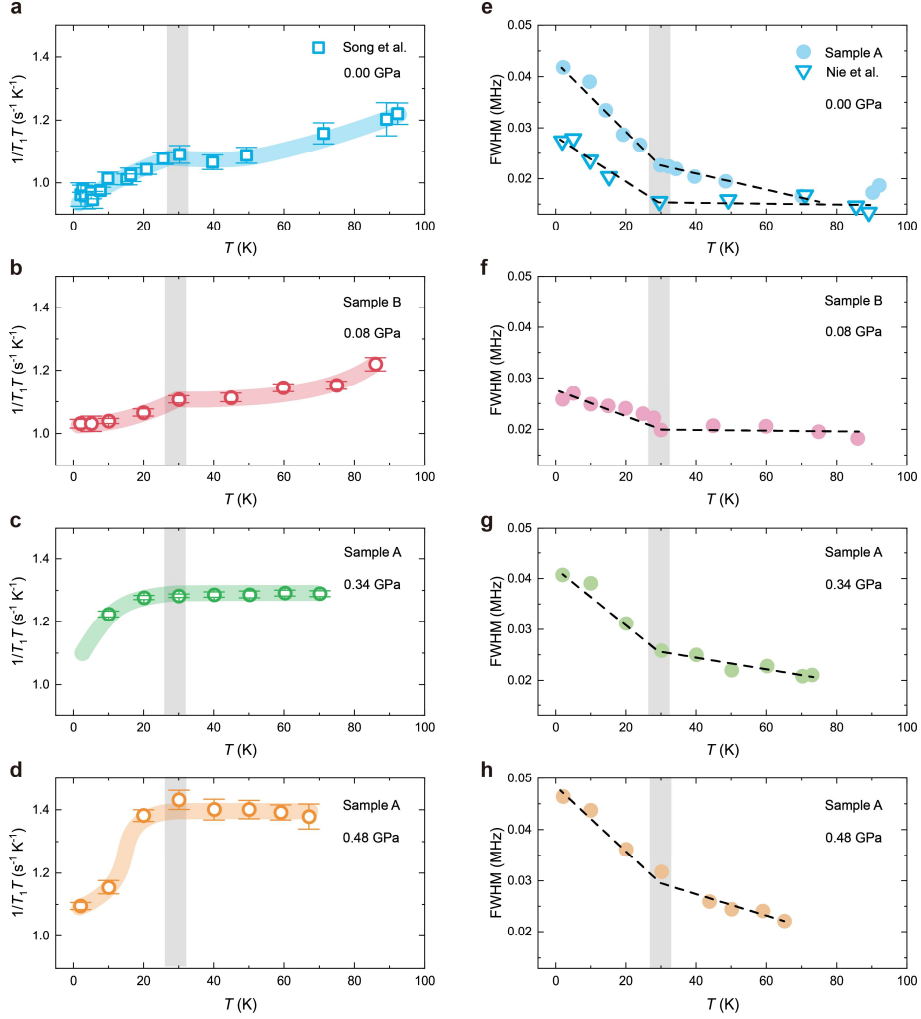

**Extended data Fig. 2 | Determination of $T_{nem}$ under different pressures. a-d,** Temperature-dependent $1/T_1T$ in the triple-$Q$ CDW phase. The $1/T_1T$ was measured on the low-frequency peak of the central transition lines (see Fig. 2a). The nematic transition leads to an additional drop in $1/T_1T$ below $T_{nem}$ [53]. With increasing pressure, $T_{nem}$ is nearly unchanged up to 0.48 GPa. The colored line is a guide for the eyes. The grey vertical line marks $T_{nem}$. **e-h,** The evolution of the full width at half maximum (FWHM) in the triple-$Q$ CDW phase. The FWHM was also measured on the low-frequency peak of the central transition lines. Similar to $1/T_1T$, a sudden increase in FWHM is also revealed below $T_{nem}$ [9]. The black dashed line is a guide for the eyes. $1/T_1T$ data for 0 GPa come from ref. [32]. One of FWHM data for 0 GPa comes from ref. [9]. Except for data for 0.08 GPa is measured on sample B, and data for other pressures are measured on sample A.

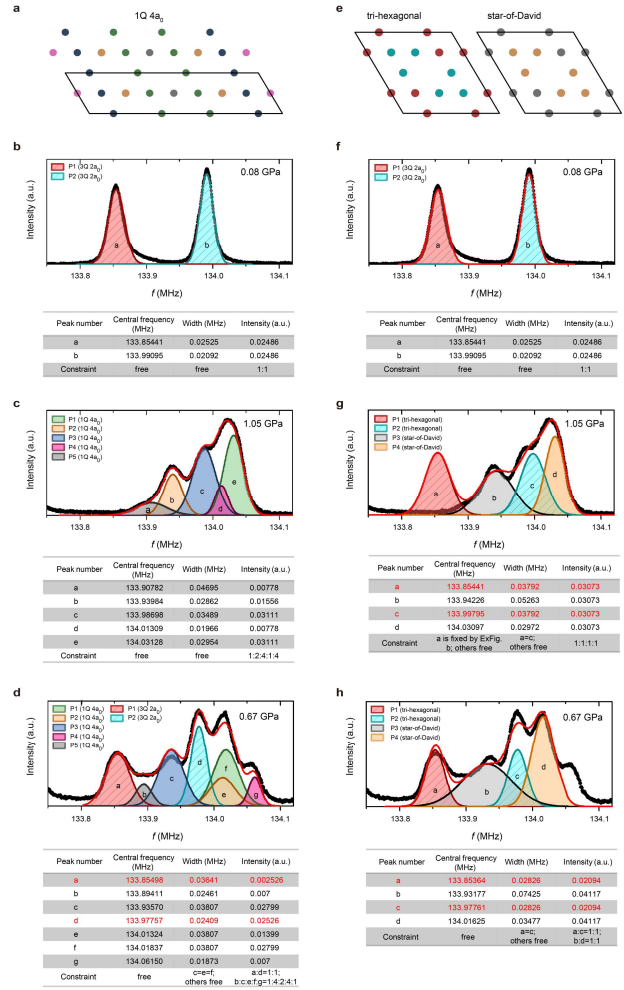

**Extended data Fig. 3 | Analysis of pressure-dependent NMR spectra with different CDW models.**
**a,** Different V sites in a stripe-like CDW order with unidirectional $4a_0$ modulation. Different colored dots represent different V sites. **b-d,** Analysis of NMR spectra with stripe-like CDW order with a unidirectional $4a_0$ modulation. Black circles are experimental data points. Red lines are the sum of all fitting curves. The two colorful peaks with the same intensity in **b** and **f** represent two $^{51}$V sites in triple-$Q$ CDW order. The five colorful peaks with an intensity ratio of 1:1:2:4:4 in **c** are the fitting result with stripe-like CDW order. The spectrum in **d** can be fitted by a combination of triple-$Q$ CDW order and stripe-like CDW order. The fitting constraint for both CDW orders in **d** is the same as that in **b** and **c**. **e,** different V sites in a superimposed tri-hexagonal and star-of-David model. **f-h,** Analysis of NMR spectra with the superimposed tri-hexagonal and star-of-David model. The NMR spectrum for triple-$Q$ CDW at ambient pressure is well explained by the tri-hexagonal (or star-of-David) model [32]. The star-of-David (or tri-hexagonal) modulation also splits the NMR spectrum into two peaks but the position should be different from that in tri-hexagonal modulation. Therefore, the superimposed tri-hexagonal and star-of-David model should exhibit two sets of double peaks in the NMR spectrum. Obviously, this kind of model cannot fit the NMR spectrum under 1.05 GPa in **g**. Moreover, the NMR spectrum in **h** cannot be fitted by a combination of two kinds of triple-$Q$ CDW orders in **f** and **g**. All data come from sample B. Each table below the figure shows the fitting parameter and fitting constraint used.

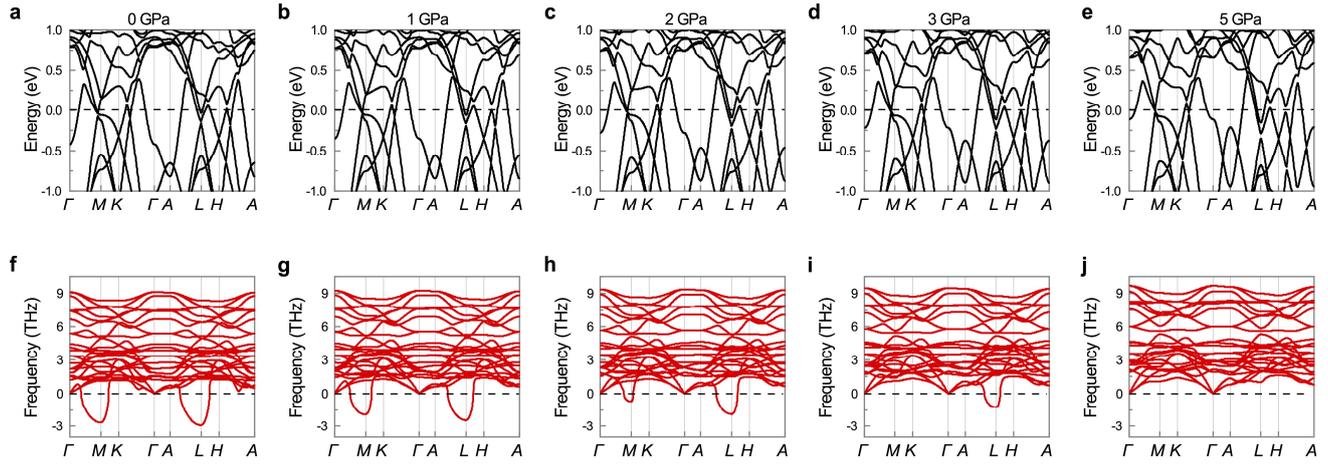

**Extended data Fig. 4 | Pressure-dependent electronic structure and phonon spectra. a-e,** Pressure-dependent DFT calculation of electronic structures without CDW modulation. Both Fermi surface topology and van Hove singularities remain nearly unchanged below 3 GPa. **f-j,** Pressure-dependent DFT calculation of phonon spectra without CDW modulation. The pressure is the same as the electronic structures on each phonon spectrum. The imaginary frequency at $M$ point disappears above 2 GPa, but the imaginary frequency at $L$ point remains at 3 GPa and finally disappears at 5 GPa. This result suggests that the CDW transition is expected up to 3 GPa, which is inconsistent with the observation in the present NMR experiment.

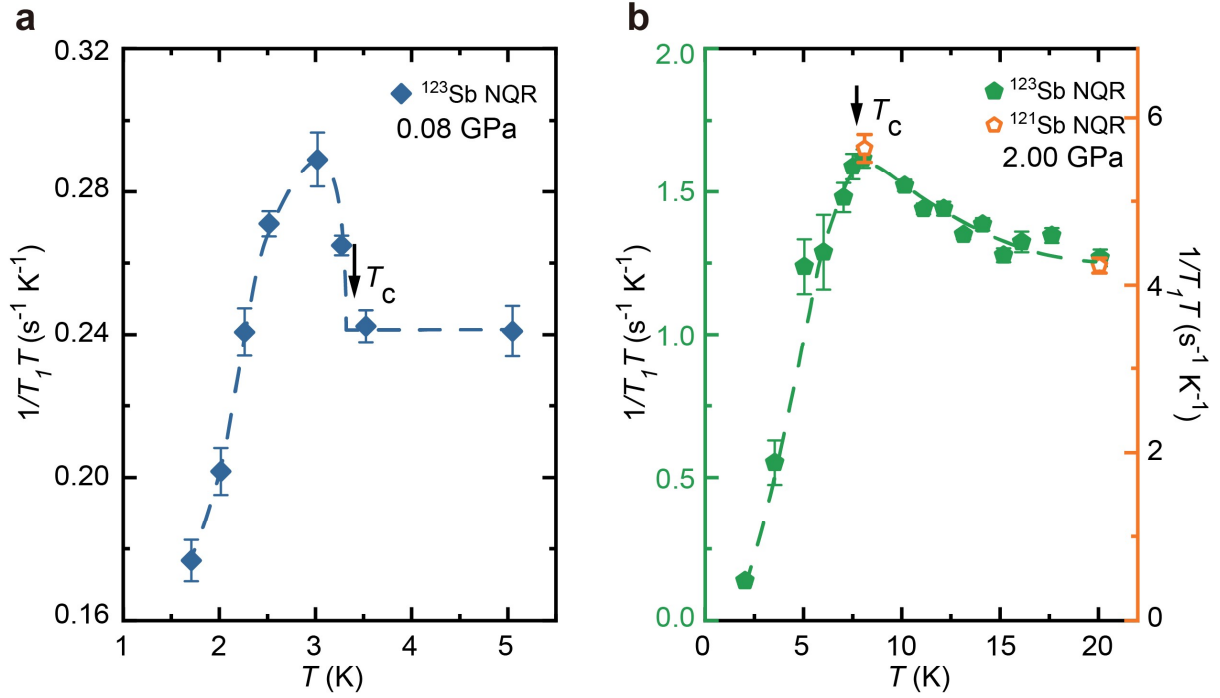

**Extended data Fig. 5 | Comparison of NQR measured $1/T_1T$ at 0.08 GPa and 2.00 GPa. a,** Low-temperature $1/T_1T$ at 0.08 GPa by [123]Sb NQR measurement on Sb1 sites. The dashed line is the guide for the eyes; **b,** temperature-dependent $1/T_1T$ at 2.00 GPa by [123]Sb and [121]Sb NQR measurements on Sb1 sites. The filled and open pentagons represent $1/T_1T$ measured on [123]Sb and [121]Sb respectively. The longitudinal axis scale on the right-hand side belonging to [121]Sb NQR is 3.41 times larger than that on the left-hand side belonging to [123]Sb NQR. Usually, the relaxation rate for the $T_1$ process in NQR has two kinds of relaxation channels. One is the magnetic relaxation channel and the other is the quadrupole relaxation channel. These two relaxation channels can be identified by checking the ratio between the relaxation rates at different isotopes. The magnetic relaxation channel requires $T_1^{-1} \sim \gamma_n^2$, where $\gamma_n$ is the nuclear gyromagnetic ratio. The expected ratio between nuclear spin-lattice relaxation rate at [121]Sb and [123]Sb through the magnetic relaxation channel is expressed as $\frac{T_{1M}^{-1}(121)}{T_{1M}^{-1}(123)} = \frac{(10.189)^2}{(5.51756)^2} = 3.41$. The quadrupole relaxation channel requires $T_1^{-1} \sim 3(2I + 3)Q^2/[10(2I - 1)I^2]$, where $Q$ is the quadrupole moment. Thus, the expected ratio between the nuclear spin-lattice relaxation rate at [121]Sb and [123]Sb through the quadrupole relaxation channel is $\frac{T_{1Q}^{-1}(121)}{T_{1Q}^{-1}(123)} = 1.5$ [65]. We checked this ratio at two temperatures above $T_c$ and found that it is very close to 3.41. Therefore, the magnetic relaxation channel dominates $T_1$ in the NQR measurement.

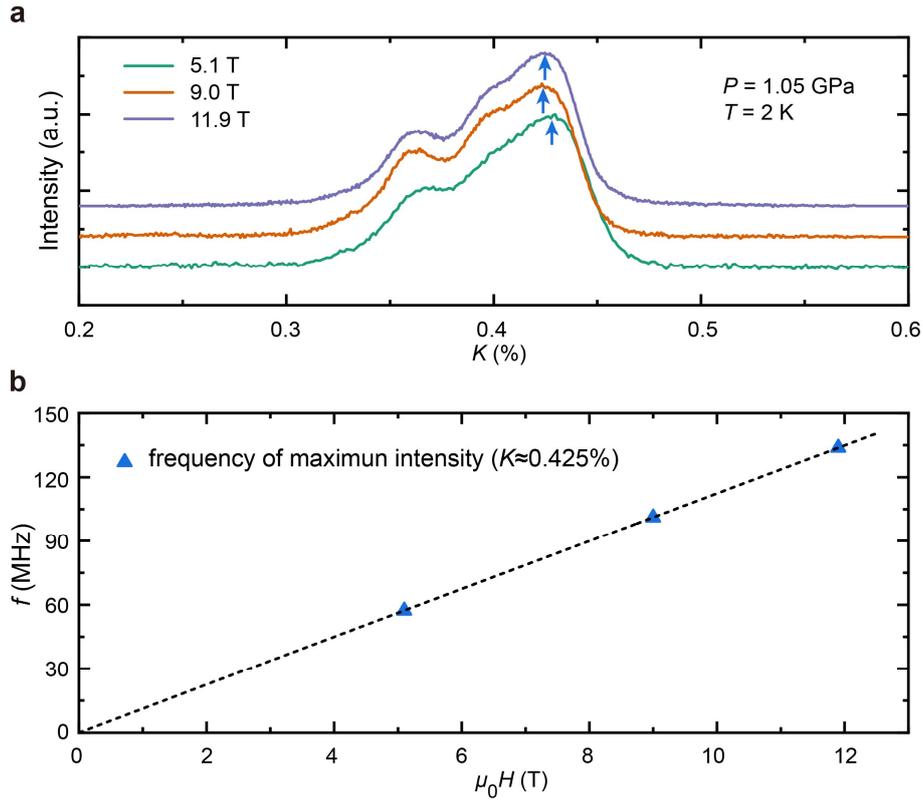

**Extended data Fig. 6 | Field dependence of the $^{51}$V NMR spectrum at 2 K. a**, Central lines of the $^{51}$V NMR spectrum in 5.1 T, 9.0 T and 11.9 T. The horizontal axis is the Knight shift. If the NMR spectrum is fully paramagnetic, the spectrum is unchanged under different applied magnetic fields. **b**, NMR frequency vs applied magnetic field. The NMR frequency is taken as the maximum intensity (marked by blue arrows in **a**). The dashed line is a linear fitting nearly across the zero point with an error bar of 0.0003 T. No significant evidence of a magnetic moment is expected here. All data points are collected in sample C.

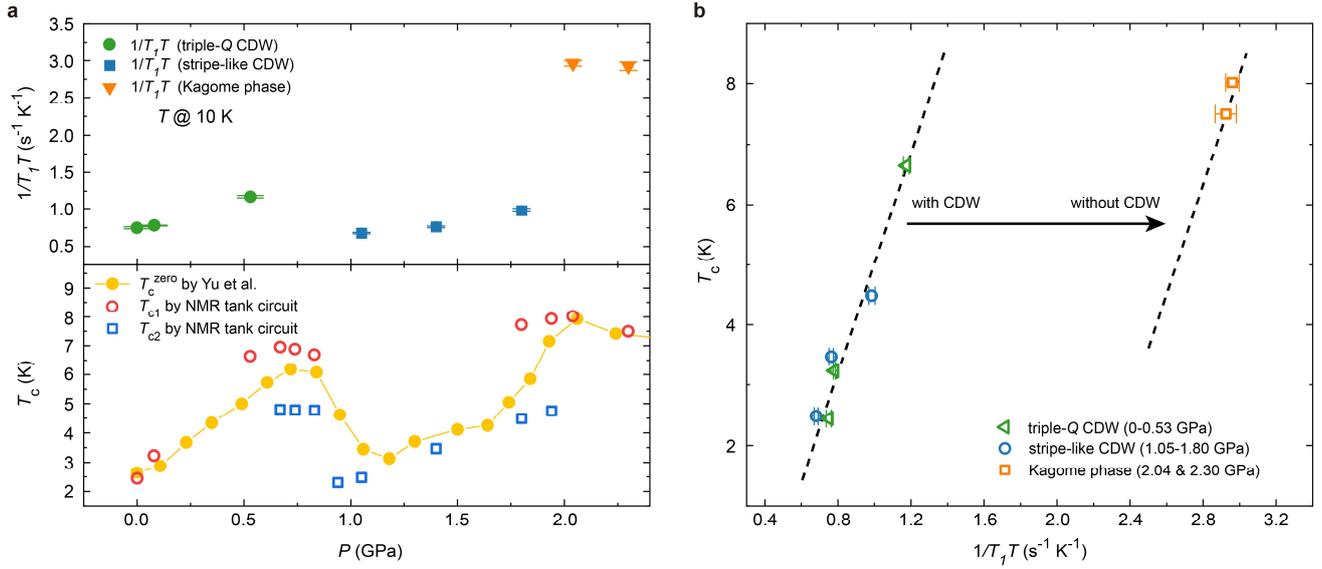

**Extended data Fig. 7 | Correlation between $T_c$ and $1/T_1T$ under pressure. a,** Pressure-dependent $1/T_1T$ and $T_c$. $1/T_1T$ under each pressure is calculated by the weighted average method that sums up $1/T_1T$ measured on each peak with its spectrum weight. The spectrum weight is determined by the normalized integration area of $1/T_1T$ as shown in Extended Fig. 9. All $1/T_1T$ data were collected at 10 K above the maximum value of $T_c$. To avoid effect from the phase transition, only $1/T_1T$ belonging to a pure phase is plotted here. All data points are collected in sample B. **b,** Plot of $1/T_1T$ vs $T_c$ under the same pressure in **a**. All data points can be separated into two groups. It is obvious that the data in different CDW phases (triple-$Q$ CDW and stripe-like CDW) exhibit a similar linear behavior, supporting a strong correlation between $T_c$ and $1/T_1T$ in the CDW phase. In the kagome phase, the data seem to follow a similar linear behavior with the same slope but the disappearance of the CDW order leads to a constant shift in $1/T_1T$.

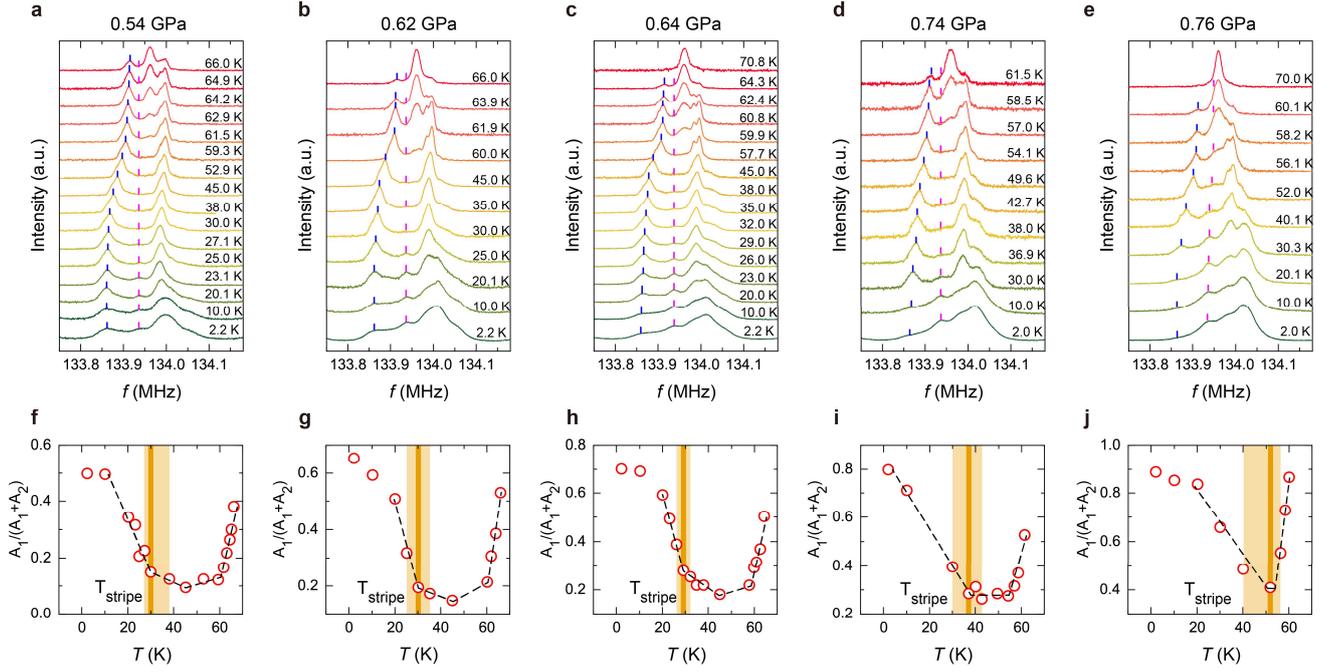

**Extended data Fig. 8 | Determination of *T*ₛₜᵣᵢₚₑ above *P*ₐ₁.** **a-e**, The temperature-dependent central lines of $^{51}$V NMR under different pressures. The spectrum configuration continuously changes from 0.54 GPa to 0.76 GPa at 2 K, which reveals the first order quantum phase transition. Meanwhile, the transition temperature of the stripe-like CDW $T_{stirpe}$ rapidly increases above 0.64 GPa and converges to $T_{CDW}$ in 0.76 GPa. **f-j**, The temperature-dependent renormalized intensity ratio of stripe-like CDW under different pressures. The renormalized intensity ratio is expressed as $A_1/(A_1+A_2)$, where $A_1$ and $A_2$ are the integral spectrum weights on the characteristic peaks belonging to triple-$Q$ CDW and stripe-like CDW respectively. The characteristic peaks of triple-$Q$ CDW and stripe-like CDW are labelled by blue and magenta bars, respectively. The spectral integration on both triple-$Q$ CDW and stripe-like CDW is performed on the corresponding peak frequency with a width of 10 kHz. All data were collected from sample A.

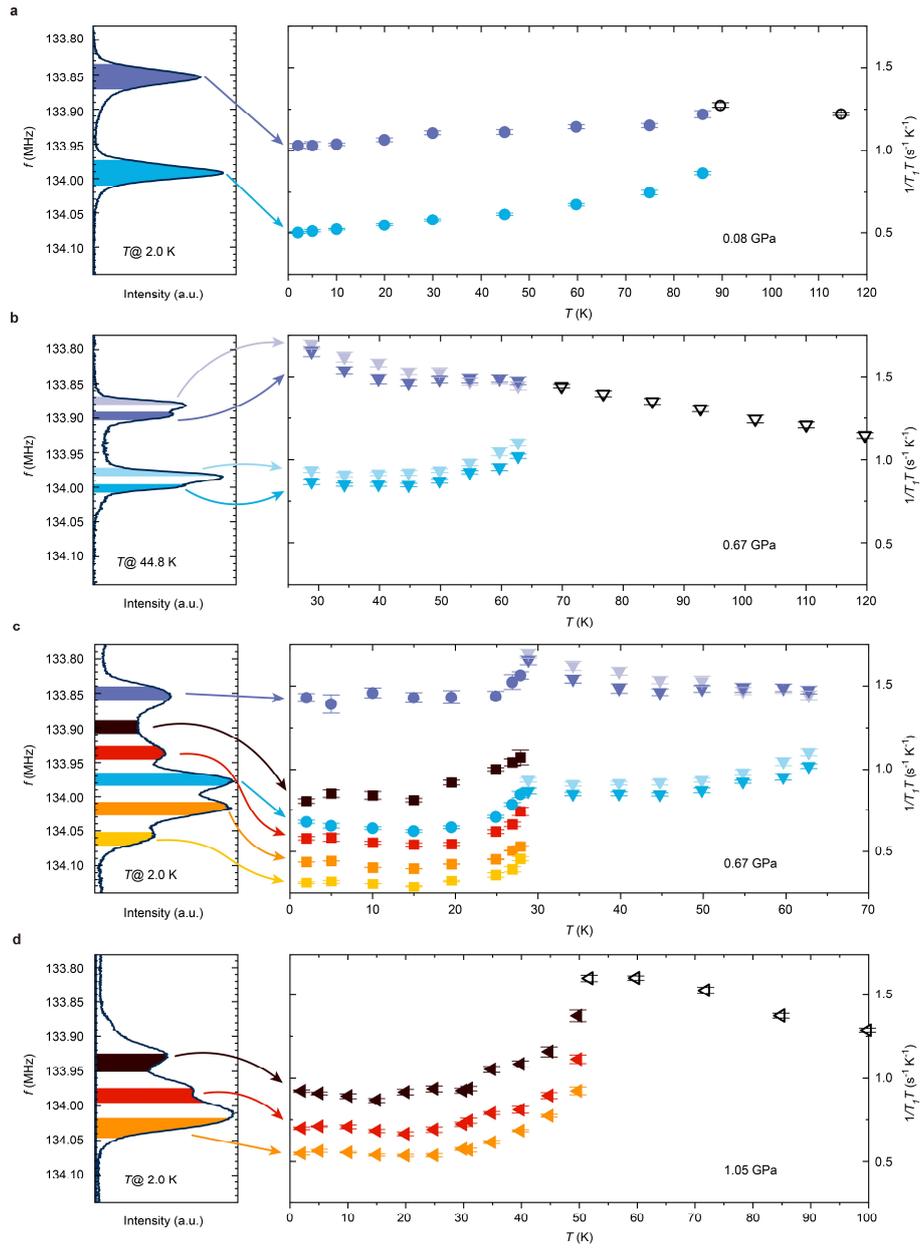

**Extended data Fig. 9 | Integral area for nuclear spin-lattice relaxation measurement in different CDW states as shown in Fig. 3.** Integral area of $1/T_1T$ analysis for **a** 0.08 GPa at 2.02 K; 0.67 GPa **b** at 44.8 K and **c** at 2.0 K; **d** 1.05 GPa at 2.0 K. Each colorful block on the left represents the frequency integral region for each value of $1/T_1T$ on the right with the same color. Here, $1/T_1T$ is the same as that in Fig. 3.

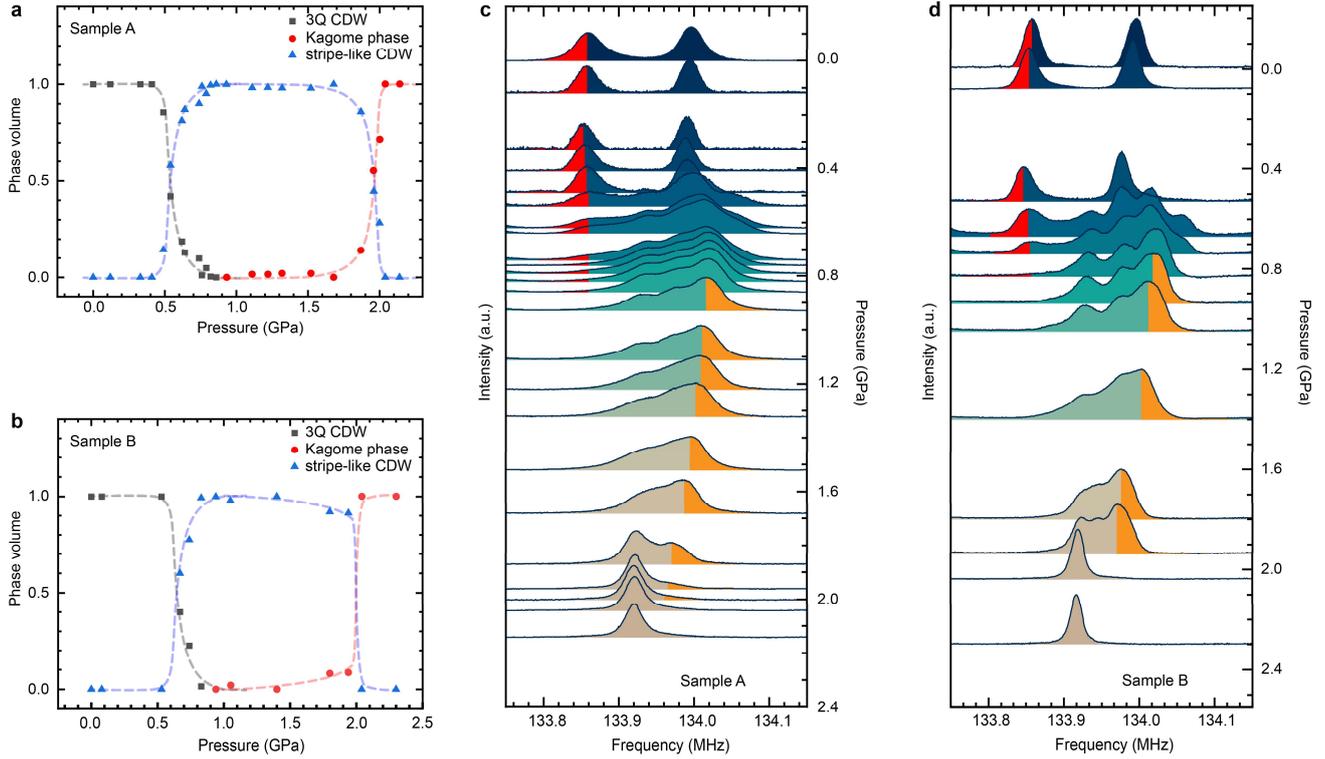

**Extended data Fig. 10 | Analysis of pressure-dependent phase volume. a-b**, Pressure dependence of the phase volume for sample A and sample B. The phase volume is determined through the weight of the characteristic peak belonging to different phases. The weight of the characteristic peak is calculated by integrating the intensity on part of the peak and then divided by the whole area of the spectrum. Finally, the phase volume is determined by renormalizing the pressure-dependent weight of each phase to unity (value from 0 to 1). Here, the phase volumes near ambient pressure and those above 2.00 GPa are manually defined as 0 or 1 since they are pure phases. The standard weight of the characteristic peak belonging to the pure stripe-like phase is estimated to be near 1.2 GPa for the two samples since the weight at approximately 1.2 GPa is nearly unchanged with pressure. The integration areas of each phase are shown in **c** and **d**. **c-d**, The integrating areas for the calculation of the phase volume. The red block is the integration area belonging to the 3Q CDW phase. The yellow block is the integration area belonging to stripe-like CDW.